\documentclass[aps,prd,twocolumn,preprintnumbers,floatfix,nofootinbib]{revtex4-2}
\usepackage[utf8]{inputenc}
\usepackage[dvips]{graphicx}
\usepackage[dvipsnames]{xcolor}
\usepackage{color}
\usepackage{relsize}
\usepackage{graphics}
\usepackage{epstopdf}
\usepackage{hyperref}
\usepackage{mathrsfs}
\usepackage{amssymb}
\usepackage{physics}
\usepackage{booktabs}
\usepackage{float}

\usepackage[normalem]{ ulem }
\usepackage{amsthm}
\usepackage{amsmath}
\usepackage{cancel}
\usepackage[caption=false]{subfig}
\usepackage{tabularx,ragged2e} 
 \newcolumntype{L}{>{\RaggedRight\arraybackslash}X}

\usepackage{hyperref}

\newcommand{\TBH}{T}
\newcommand{\MBH}{M}
\newcommand{\MBHi}{M_{\rm in}}

\newcommand{\as}{a_\star}

\newcommand{\GeV}{\rm GeV}
\newcommand{\g}{\rm g}

\newcommand{\Tform}{T_{\rm f}}

\newcommand{\rhorad}{\rho_{\rm SM}}

\definecolor{palatd}{RGB}{104, 36, 109}
\definecolor{palatb}{RGB}{0, 56, 168}
\definecolor{palatr}{rgb}{0.745,0.118,0.176}
\newcommand\myshade{80}
\colorlet{mylinkcolor}{palatr}
\colorlet{mycitecolor}{palatb}
\colorlet{myurlcolor}{palatd}

\hypersetup{
  linkcolor  = mylinkcolor!\myshade!black,
  citecolor  = mycitecolor!\myshade!black,
  urlcolor   = myurlcolor!\myshade!black,
  colorlinks = true
}

\usepackage{pgfplots}

\pgfplotsset{compat=1.17}

\begin{document}
\sloppy  

\preprint{IFT-UAM/CSIC-25-11}

\title{Page Time of Primordial Black Holes in the Standard Model and Beyond}

\author{Yuber F. Perez-Gonzalez}
\email{yuber.perez@uam.es}
\affiliation{Departamento de F\'{i}sica Te\'{o}rica and Instituto de F\'{i}sica Te\'{o}rica (IFT) UAM/CSIC, Universidad Aut\'{o}noma de Madrid, Cantoblanco, 28049 Madrid, Spain}

\begin{abstract}
    The Page time marks the moment when the von Neumann entropy of the emitted Hawking radiation equals the Bekenstein-Hawking entropy of an evaporating black hole, which is assumed to quantify its degrees of freedom as seen from the outside.
    Beyond this point, from unitarity we would expect that the entropy of the radiation begins to decrease, ensuring that information is eventually recovered.
    In this work, we investigate the dependence of the Page time on black hole properties and the particle content of nature. 
    Specifically, we analyze its sensitivity to the Standard Model (SM) and potential Beyond-the-SM degrees of freedom, incorporating the effects of particle masses.
    We find that a Schwarzschild primordial black hole (PBH) with an initial mass of $\sim 6.23\times 10^{14}~{\rm g}$ would have a Page time equal to the age of the Universe, assuming emission of SM particles only.
    We further explore the impact of a non-negligible PBH angular momentum, finding that light spin-2 particles are predominantly emitted before the Page time. 
    Specifically, for initial angular momenta values exceeding $a_\star > 0.5$, approximately $70\%$ of the total graviton emission occurs prior to the Page time for PBHs with an initial mass $M_{\rm BH} \lesssim 10^{10}~{\rm g}$. 
    Finally, we discuss the implications for PBH phenomenology, particularly regarding potential constraints from $\Delta N_{\rm eff}$ measurements.
\end{abstract}

\maketitle

\section{Introduction}

Quantum field theory in curved spacetime predicts that black holes (BHs) possess a temperature and emit radiation, a phenomenon known as Hawking radiation~\cite{Hawking:1974rv,Hawking:1974sw}. 
This discovery completed the thermodynamic description of BHs, expanding upon Bekenstein's pioneering work~\cite{Bekenstein:1972tm}, which established that a black hole's entropy is proportional to the area of its event horizon. 
The resulting Bekenstein-Hawking entropy, a thermodynamic quantity, suggests that black holes may be quantum objects with a finite number of internal degrees of freedom.

However, a fundamental issue arises with this description.  
Consider a BH formed from the collapse of a system initially in a pure state.  
At early stages, the emitted Hawking radiation appears thermal and thus resembles a mixed state.  
Note that this does not pose a problem.  
In the standard picture of Hawking evaporation, radiation emerges from the splitting of particle-antiparticle pairs, where one particle escapes to infinity while the other falls into the black hole interior.  
Since these pairs are entangled, the accessible Hawking radiation is naturally in a mixed state.  
In other words, the radiation remains entangled with the black hole, ensuring that the combined system remains in a pure state.

The issue arises when considering the evolution of entropy.  
In the semiclassical framework, the von Neumann entropy of the radiation continues to grow throughout the evaporation process, while the Bekenstein-Hawking entropy steadily decreases.  
At a specific time, known as the Page time, the two entropies become equal.  
If the Bekenstein-Hawking entropy quantifies the BH's degrees of freedom from the perspective of an outside observer, as postulated by the \emph{central dogma}~\cite{Almheiri:2020cfm}, then beyond the Page time, the radiation entropy surpasses the black hole's available degrees of freedom.
At this stage, the radiation can no longer remain fully entangled with the black hole, leading to an apparent transition from a pure state to a mixed one.  
This violates unitarity and gives rise to the \emph{information paradox}~\cite{Hawking:1976ra,Strominger:1994tn,Mathur:2009hf,Almheiri:2020cfm,Buoninfante:2021ijy,Buoninfante:2024oxl}.

To resolve this paradox, the entropy of radiation must begin to decrease after the Page time, following the widely known Page curve~\cite{Page:1993wv,Page:2004xp,Page:2013dx}.  
Recently, using the gravitational path integral, the Page curve has been derived by applying the Ryu-Takayanagi formula~\cite{Ryu:2006bv} within the framework of what is referred to as the ``island program''~\cite{Penington:2019npb,Penington:2019kki,Almheiri:2019psf,Almheiri:2019hni,Almheiri:2019qdq}.  
Although this approach raises several open questions~\cite{Geng:2021hlu,Omiya:2021olc,Harlow:2020bee}, it has provided crucial perspectives on BH evolution beyond the Page time.  
Nonetheless, various alternative resolutions to the information paradox have been proposed~\cite{Preskill:1992tc,Mathur:2009hf,Hawking:2016msc,Ho:2024tby,Pegas:2024abf}, along with debates on the validity of the central dogma, highlighting both supporting and opposing perspectives~\cite{Strominger:1996sh,Ashtekar:1997yu,Rovelli:1996dv,Meissner:2004ju,Dvali:2011aa,Dvali:2013eja,Buoninfante:2021ijy}. Additionally, some arguments challenge the very existence of the paradox itself~\cite{Buoninfante:2021ijy}.

Most studies addressing the information paradox have focused on its theoretical aspects or proposed resolutions.  
This naturally raises the question of whether experimental insights could be gained from direct observations of black holes evaporating beyond the Page time, as well as the potential implications of the paradox for phenomenology.  
To directly observe Hawking evaporation, we would need black holes with masses much smaller than those formed through stellar collapse. 
One possible candidate are primordial black holes (PBHs), which may have formed in the Early Universe.

The possible existence of PBHs and their influence on cosmic evolution have garnered significant interest, particularly following the detection of gravitational waves (GWs) and the confirmation of astrophysical black holes~\cite{LIGOScientific:2016aoc}.  
Depending on their initial mass, strong constraints exist on the abundance of PBHs that could have existed or still persist in the Universe~\cite{Carr:2020gox,Carr:2020xqk,Khlopov:2008qy}.  
A PBH with an initial masses below a few times $10^{14}~\g$ would have either completely evaporated or be undergoing its final stages of evaporation today, assuming that only Standard Model (SM) degrees of freedom are involved.  
As a result, constraints on PBHs in this mass range inherently rely on the assumption that the semiclassical framework remains valid beyond the Page time.  

Beyond their theoretical significance, PBHs have been extensively studied as potential sources of new physics.  
Various works have explored the possibility that PBHs produced via Hawking evaporation the observed Dark Matter~\cite{Lennon:2017tqq,Morrison:2018xla,Hooper:2019gtx,Auffinger:2020afu,Gondolo:2020uqv,Bernal:2020bjf,Bernal:2020ili,Bernal:2020kse,Baldes:2020nuv,Masina:2020xhk,Masina:2021zpu,Sandick:2021gew,Bernal:2021bbv,Bernal:2021yyb,Cheek:2021odj,Cheek:2021cfe,Barman:2021ost,Bernal:2022oha,Cheek:2022mmy,Chen:2023lnj,Chen:2023tzd,Kim:2023ixo,Haque:2023awl,RiajulHaque:2023cqe,Chaudhuri:2023aiv,Gehrman:2023esa,Gehrman:2023qjn,Carr:2016drx,Clesse:2016vqa,Carr:2017jsz,Green:2020jor,Croon:2020ouk},  
contributed to the production of Dark Radiation~\cite{Hooper:2019gtx,Lunardini:2019zob,Masina:2020xhk,Masina:2021zpu,Hooper:2020evu,Arbey:2021ysg,Cheek:2022dbx,Papanikolaou:2023oxq},  
or played a role in baryogenesis, either directly~\cite{Barrow:1990he,Majumdar:1995yr,Upadhyay:1999vk,Dolgov:2000ht,Bugaev:2001xr,Baumann:2007yr,Hooper:2020otu,Gehrman:2022imk}  
or via leptogenesis~\cite{Fujita:2014hha,Hamada:2016jnq,Hooper:2020otu,Perez-Gonzalez:2020vnz,Bernal:2022pue,JyotiDas:2021shi,Calabrese:2023key,Calabrese:2023bxz,Schmitz:2023pfy,Ghoshal:2023fno,Datta:2020bht,Gunn:2024xaq,Calabrese:2025sfh}.  
Additionally, PBHs have been linked to gravitational wave generation~\cite{Papanikolaou:2020qtd,Domenech:2020ssp,Bhaumik:2022pil,Bhaumik:2022zdd,Papanikolaou:2022chm,Ghoshal:2023sfa}  
and the stability of the SM Higgs potential~\cite{Burda:2016mou,Burda:2015isa,Hamaide:2023ayu}.  
Given these broad implications, it is natural to ask whether modifications to Hawking radiation beyond the Page time could significantly alter the conclusions drawn from these phenomenological studies.

In this work, we take an initial step in this direction by analyzing the dependence of the Page time and Page curve on PBH properties, such as mass and angular momentum, as well as on the particle content of nature, considering both the SM and beyond-the-SM scenarios.  
More broadly, this work aims to introduce the entropy problem to the phenomenology community in an accessible manner for those unfamiliar with its implications.  
Although the modifications to the Hawking spectrum beyond the Page time remain uncertain, we outline potential effects on the phenomenological studies mentioned above.  
As a benchmark, we examine the generation of dark radiation in the form of gravitons from a PBH population and discuss possible deviations from standard predictions.  

This paper is organized as follows.  
In Sec.~\ref{sec:BH_evap}, we review the general properties of Kerr BH evaporation and describe their evolution within the semiclassical framework.  
Sec.~\ref{sec:Entrop_Pt} introduces the different definitions of entropy relevant to this discussion, including the Bekenstein-Hawking thermodynamic entropy and the von Neumann entropy of Hawking radiation.  
We also examine how the Page time and Page curve depend on BH properties and the particle spectrum.  
To connect with phenomenology, Sec.~\ref{sec:PBHs} discusses the formation and evolution of PBHs, the observational constraints on their abundance, and the implications of the Page time in this context.  
Additionally, we explore how modifications to BH entropy evolution beyond the Page time could impact existing constraints and phenomenological results.  
Finally, Sec.~\ref{sec:Conc} presents concluding remarks.  
Throughout this work, we adopt natural units where $\hbar = c = k_{\rm B} = 1$.

\section{Black Hole Evaporation}\label{sec:BH_evap} 

Consider a Kerr black hole, characterized by its instantaneous mass $\MBH$ and dimensionless spin parameter $\as \equiv J /(G\MBH^2) \in [0, 1)$, where $J$ represents the black hole’s angular momentum. 
In Boyer-Lindquist coordinates $(t,r,\theta,\phi)$, its spacetime metric is given by~\cite{Boyer:1966qh}
\begin{widetext}
\begin{align}\label{eq:BL_metric}
    \dd s^2=\frac{\Delta}{\Sigma}\,(\dd t - a \sin^2\theta \dd\phi)^2 - \frac{\sin^2\theta}{\Sigma}\,(-a \dd t +(r^2+a^2)\dd\phi)^2-\frac{\Sigma}{\Delta}\,\dd r^2 - \Sigma\, \dd\theta^2,
\end{align}
\end{widetext}
Here, $\Delta \equiv r^2 - 2 G\MBH r + a^2$ and $\Sigma \equiv r^2 + a^2\cos^2\theta$, with $a = \as G\MBH$. 
Two key quantities characterize the black hole’s thermodynamics.
First, the area of the outer horizon is
\begin{align}
    A = 8\pi G^2 M^2 \lambda(a_\star),
\end{align}
where, for compactness, we define
\begin{align}
    \lambda(a_\star) \equiv 1+\sqrt{1-a_\star^2},
\end{align}
such that $\lambda(0) = 2$ for Schwarzschild black holes. 
The surface gravity at the outer horizon, which determines the black hole's temperature, is given by
\begin{align}
    \kappa_+ = \frac{r_+ - r_-}{2(r_+^2 + a^2)},
\end{align}
where $r_\pm$ are the radii of the event horizons
\begin{align}
    r_\pm = GM(1 \pm \sqrt{1-a_\star^2}).
\end{align}
Assuming an initial vacuum state in the “in” basis at past infinity, Hawking showed that an observer at future infinity measures a nonzero expectation value of the number operator~\cite{Hawking:1974rv,Hawking:1974sw}.
This value, evaluated in the ``out'' basis---corresponding to the basis of the observer at future infinity---matches that of a body emitting thermal radiation. The emission rate for a particle species $i$ over an interval of time ($\dd t$) and energy ($\dd E$) is given by
\begin{align}\label{eq:KBHrate}
\frac{\dd^2 \mathcal{N}_{i}}{\dd E\dd t}&=\frac{g_i}{2\pi} \sum_{l=s_i}\sum_{m=-l}^l \mathcal{N}_{ilm}\,,
\end{align}
where the rate per mode $\mathcal{N}_{ilm}$ is
\begin{align}\label{eq:KBHrate_lm}
\mathcal{N}_{ilm} &=\frac{\Gamma_{s_i}^{lm}(x,a_\star)}{\exp\left[(E_i - m \Omega)/\TBH\right]-\epsilon_i}\,,
\end{align}
with $E_i$ ($g_i$) being the energy (internal degrees of freedom) of particle $i$, and $l, m$ representing the total and axial angular momentum quantum numbers, respectively. 
Here, $\epsilon_i\equiv (-1)^{2s_i}$ depends on the spin $s_i$ of the particle. The black hole temperature $T$ is 
\begin{align}
    T = \frac{\kappa_+}{2\pi} = \frac{1}{4\pi G\MBH} \frac{\sqrt{1-\as^2}}{\lambda(a_\star)},
\end{align}
being $\Omega = \as/(2GM\lambda(\as))$ the horizon's angular velocity.
In Eq.~\eqref{eq:KBHrate_lm}, $\Gamma_{s_i}^{lm}$ is the greybody factor, which quantifies the probability that an emitted particle escapes to infinity rather than being reabsorbed by the black hole~\cite{Hawking:1974rv,Hawking:1974sw,Page:1976df,Page:1977um}. These probabilities are determined by calculating the reflection and transmission coefficients for scattering processes involving the gravitational potential.
Such coefficients are obtained by solving the equations of motion in the curved spacetime surrounding the black hole.
Importantly, the equations of motion for a spacetime described by the line element in Eq.~\eqref{eq:BL_metric} are separable, leading to the well-known Teukolsky master equations for the radial and angular components~\cite{Teukolsky:1973ha,Press:1973zz,Teukolsky:1974yv}.
We have numerically solved these Teukolsky equations for particles with integer spins, following the procedure established in Ref.~\cite{Dong:2015yjs}, taking the relevant coefficients for $s=0,1,2$ from Ref.~\cite{Teukolsky:1974yv}.
For fermions, we have applied the variable transformation of Ref.~\cite{Page:1977um}, and solved the system of differential equations presented in the same reference.
We have verified that we recover the absorption probabilities appearing in the literature for all cases.

The emission rate in Eq.~\eqref{eq:KBHrate_lm} corresponds to the expectation value for the emission of a single particle $i$ in a quantum state defined by the numbers $l, m$, and energy $E_i$, i.e.,~$\langle n_{ilm} \rangle = \mathcal{N}_{ilm}$. By following a similar approach, it is possible to compute the expectation values for the emission of $p$ particles in the same state, $\langle n_{ilm}^p \rangle$, where $p \in [0,1]$ for fermions and $p \in \mathbb{N}_0$ for bosons.
From these, we can determine the probability distribution, $P_{p}^{ilm}$, for $p$ particles in the $i,l,m$ state, cf.~Refs.~\cite{Wald:1975kc,Parker:1975jm,Hawking:1976ra}
\begin{align}
    P_{p}^{ilm} =  \,\mathcal{N}_{ilm}^p\,[1 + \epsilon_i\, \mathcal{N}_{ilm}]^{-p - \epsilon_i},
\end{align}
Building on this, one can examine whether there are correlations between the probabilities of emitting different numbers of particles in the same or distinct modes. In Refs.~\cite{Wald:1975kc,Parker:1975jm,Hawking:1976ra}, it was shown that all such correlations vanish, establishing that particle emission from black holes is purely thermal. 
As a result, the density matrix $\rho_{pq}^{ilm}$ for the Hawking radiation of a particle species $i$ corresponds to that of a system emitting thermal radiation. Specifically, it is diagonal in the late time ``out" basis with well-defined particle numbers~\cite{Wald:1975kc,Parker:1975jm,Hawking:1976ra}
\begin{align}\label{eq:den_mat}
    \rho_{pq}^{ilm} = \delta_{pq}\,P_{p}^{ilm}.
\end{align}
Thus, knowing the density matrix, we can compute the expectation value of any observable associated to the Hawking radiation.

During evaporation, BHs emit particles of various kinds at rates that depend on its instantaneous mass and angular momentum as well as on the particle's properties. 
Thus, the BH mass and angular momentum decrease at rates that can be determined by the initial BH properties and the existing degrees of freedom in nature. The mass and spin loss rates are computed by summing Eq.~\eqref{eq:KBHrate} over the different particle species and integrating over the phase space, as shown in \cite{PhysRevD.41.3052,PhysRevD.44.376}
\begin{subequations}\label{eq:dynamicsevaporation}
\begin{align}
 \frac{d\MBH}{dt} &= - \varepsilon(\MBH, a_\star)\frac{1}{G^2\MBH^2}\,,\\
 \label{eq:dynamicsspin}\frac{da_\star}{dt} &= - a_\star[\gamma(\MBH, a_\star) - 2\varepsilon(\MBH, a_\star)]\frac{1}{G^2\MBH^3}\,,
\end{align}
\end{subequations}
where $\varepsilon(\MBH, a_\star) = \sum_i \varepsilon_i(\MBH,a_\star)$ and $\gamma(\MBH, a_\star) = \sum_i \gamma_i(\MBH,a_\star)$ are the mass and angular-momentum evaporation functions, respectively.
For a given particle type $i$, the functions $\gamma_i(\MBH,a_\star)$ and $\varepsilon_i(\MBH,a_\star)$ are obtained via
\begin{subequations}\label{eq:evap_func}
\begin{align}
 \varepsilon_i(\MBH, a_\star) &= \frac{g_i}{2\pi}\int_{z_i}^\infty \sum_{l=s_i}\sum_{m=-l}^l\, x\,\mathcal{N}_{ilm}\,  dx\,,\\
 \gamma_i(\MBH, a_\star) &= \frac{g_i}{2\pi}\int_{z_i}^\infty \sum_{l=s_i}\sum_{m=-l}^l\, \frac{m}{a_\star}\, \mathcal{N}_{ilm} \, dx\,,
\end{align}
\end{subequations}
where $x = GME$ and $z_i = GMm_i$, $m_i$ the particle's mass. 
The integration in Eqs.~\eqref{eq:evap_func} is subject to a lower bound, $E \geq m_i$, ensuring that only kinematically allowed emissions contribute. The presence of massive degrees of freedom modifies the absorption probabilities and alters the integration limits accordingly\footnote{We neglect the mass dependence in the $\Gamma_{s_i}^{lm}$ factors ---using it only as a cutoff in the integration range--- which overestimates the evaporation functions. For instance, using absorption probabilities for massive fermions around Schwarzschild black holes from Refs.~\cite{Unruh:1976fm, Doran:2005vm, Lunardini:2019zob}, the exact value at $z=0.2$ is $\varepsilon_i=4.066\times10^{-6}$, while the massless approximation with a cutoff yields $\varepsilon_i=1.741\times10^{-5}$, about 4.28 times larger.
}

Thus far, our discussion has assumed a fixed black hole background. However, Hawking evaporation gradually alters the black hole geometry, a process known as backreaction.
In the semiclassical framework adopted here, incorporating backreaction requires solving the Einstein equations with the expectation value of the regularized energy-momentum tensor, $\langle T_{\mu\nu} \rangle$, computed in the ``out'' vacuum~\cite{Bardeen:1981zz,Piran:1993tq,Strominger:1994tn,Parentani:1994ij,Massar:1994iy,Brout:1995rd,Buoninfante:2024oxl}.  
This is a highly nontrivial task, as it involves obtaining a renormalized $\langle T_{\mu\nu} \rangle$, which depends on the mode solutions of the quantum field in the evolving background.  
Simultaneously, the Einstein equations depend on the metric $g_{\mu\nu}$, its derivatives, and $\langle T_{\mu\nu} \rangle$, making the problem inherently nonlinear.  
Nevertheless, progress has been made under simplifying assumptions.  
For instance, backreaction effects have been analyzed in $1+1$ dimensions~\cite{Piran:1993tq,Strominger:1994tn}, by assuming a specific form for the energy-momentum tensor component relevant to BH mass loss~\cite{Brout:1995rd}, or in $3+1$ dimensions by approximating the renormalized $\langle T_{\mu\nu} \rangle$ as similar to its form in the absence of backreaction~\cite{Bardeen:1981zz,Massar:1994iy}. 
Additionally, numerical studies in a $3+1$ dimensional model have been conducted under the assumption that the renormalized energy-momentum tensor retains a simple structure~\cite{Parentani:1994ij}. 
Across all these approaches, a common result is that the BH mass loss rate follows the relation $dM/dt \propto -M^{-2}$.  
Note, however, that some models suggest that backreaction alters BH evolution after a certain time, see, e.g., \cite{Dvali:2011aa,Dvali:2013eja,Dvali:2020wft,Dvali:2024hsb,Basumatary:2024uwo}.
Although a complete determination of the backreaction remains an open challenge~\cite{Buoninfante:2024oxl}, in this work, we assume that the BH mass and spin evolution equations are given by Eqs.~\eqref{eq:dynamicsevaporation}.

\section{Entropy production and Page time}\label{sec:Entrop_Pt}

\begin{figure*}[t!]
    \centering
        \includegraphics[width=\textwidth]{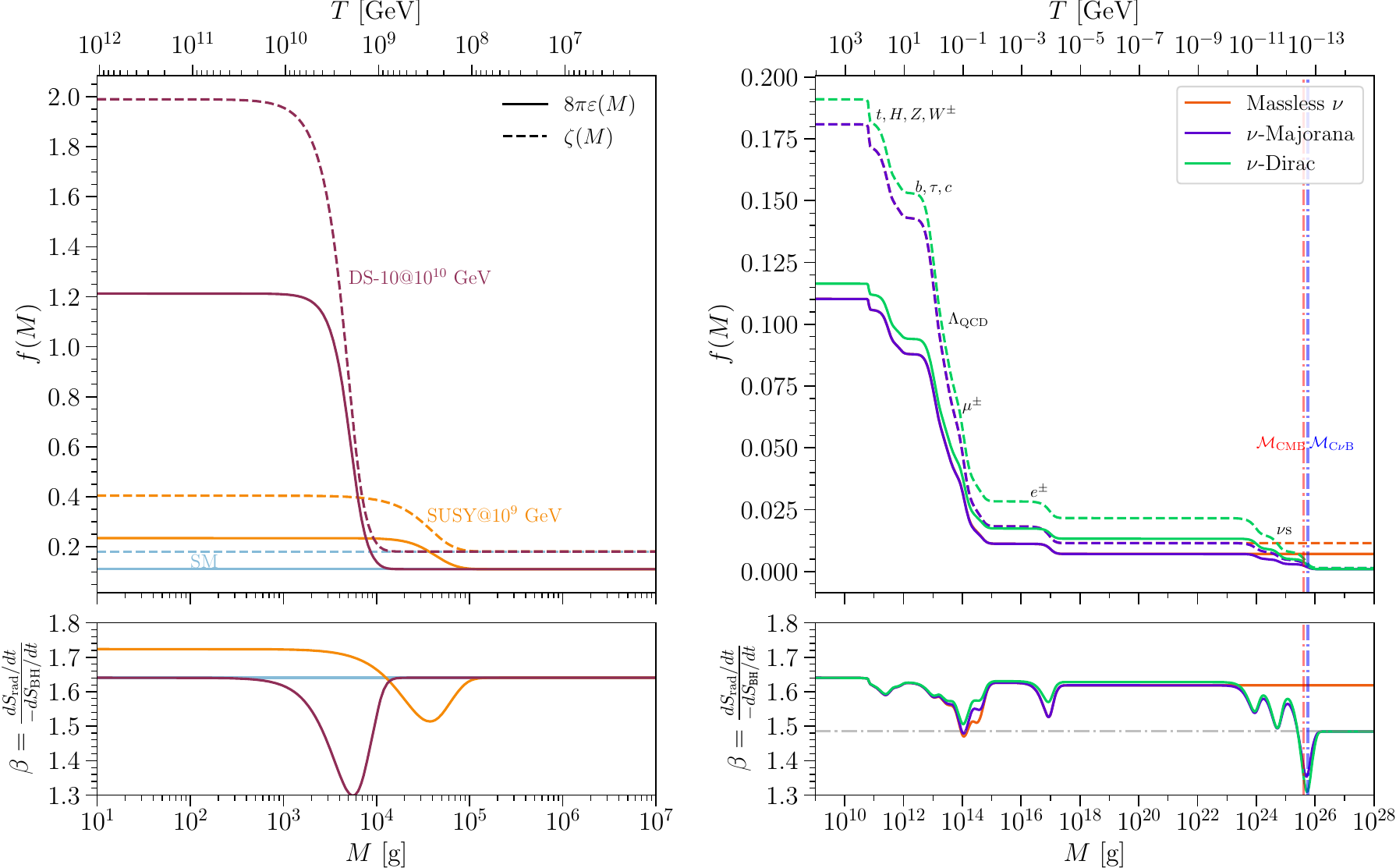}
    \caption{Top panels: Bekenstein-Hawking entropy decrease rate $8\pi\varepsilon(M)$ (solid) and von Neumann radiation entropy generation rate $\zeta(M)$ (dashed) as functions of mass for Schwarzschild BHs. On the left we consider the mass range of $10~{\rm g} \leq M \leq 10^7~{\rm g}$ for different particle sets, including the SM (light blue), the SM plus a dark sector containing ten copies of it at a scale of $10^{10}~\GeV$ (red), and the minimal supersymmetric extension of the SM at a scale of $10^9~\GeV$. On the right, we show the SM (orange), the SM plus massive Majorana neutrinos (purple), and the SM plus massive Dirac neutrinos (green) for the mass range of $10^9~{\rm g} \leq M \leq 10^{28}~{\rm g}$. The dashed vertical lines on the right panel correspond to the BH masses where the temperature equals the Cosmic Microwave Background $\mathcal{M}_{\rm CMB}$ (red) and the Cosmic Neutrino Background $\mathcal{M}_{\rm C\nu B}$ (blue), cf Ref.~\cite{Lunardini:2019zob}. Bottom panels: Entropy ratios for the same mass ranges and particle sets as in the top panels.}
    \label{fig:Page_ratio}
\end{figure*}
Having discussed the general aspects of Kerr BH evaporation, we now analyze the entropy evolution of the combined BH and emitted radiation system. 
The development of black hole thermodynamics was motivated by the discovery that the area of a black hole’s event horizon never decreases~\cite{Hawking:1971tu}.
This insight established the connection between horizon area and entropy~\cite{Bekenstein:1972tm,Bekenstein:1973ur}, leading to the formulation of the Bekenstein-Hawking entropy
\begin{align}
    S_{\tt BH} = \frac{A}{4G} =  2\pi G M^2\lambda(a_\star).
\end{align}
Additionally, the relationship between changes in the BH’s area, angular momentum, and mass closely mirrors the first law of thermodynamics:  
\begin{align*}
    dM = \frac{\kappa}{8\pi G}dA + \Omega dJ.
\end{align*}
These insights culminated in the establishment of the laws of BH thermodynamics in Ref.~\cite{Bardeen:1973gs}. Initially, the analogy between BH thermodynamics and traditional thermodynamics was considered a mathematical curiosity. However, this perspective changed significantly with the discovery of Hawking radiation and the determination of BH temperature.
More importantly, as BHs evaporate via Hawking radiation, their horizon area shrinks, necessitating a generalization of the second law of thermodynamics. This generalization incorporates the entropy of the surrounding environment $S_{\rm env}$ into the total entropy, forming the generalized second law (GSL)~\cite{Zurek:1985gd,Frolov:1993fy}
\begin{align}
    S_{\rm GSL} = S_{\tt BH} + S_{\rm env},
\end{align}
ensuring that $dS_{\rm GSL} \geq 0$.
In the semiclassical approximation, the entropy of the environment corresponds to the von Neumann entropy of the quantum fields outside the black hole~\cite{Almheiri:2020cfm}.

Next, we compute the time evolution of these two components of the GSL entropy within the same semiclassical framework.
Using the evaporation equations in Eqs.~\eqref{eq:dynamicsevaporation}, we can obtain the time evolution of the Bekenstein-Hawking entropy
\begin{align}
    \frac{d S_{\tt BH}}{dt} &= \frac{1}{4G} \frac{dA}{dt},\notag\\
    &= -2\pi \frac{[2\varepsilon \lambda(a_\star) - a_\star^2 \gamma]}{\sqrt{1-a_\star^2}} \frac{1}{G \MBH} .
\end{align}
Now, to compute, in the semi-classical approximation, the von Neumann entropy for the emitted radiation, we make use of the density matrix defined in Eq.~\eqref{eq:den_mat}.
As established in Refs.~\cite{Zurek:1982zz,Page:1983ug,Page:2004xp,Page:2013dx}, the rate of entropy generation due to radiation emission is
\begin{widetext}
\begin{align}
    \frac{d S_{\rm rad}}{dt}  &= -\sum_{i=\text{all degrees of freedom}}\int_{z_i}^\infty \sum_{l=s_i}\sum_{m=-l}^l {\rm Tr}_p[\rho^{ilm} \ln \rho^{ilm}]\notag\\
    &\equiv \zeta (M, a_\star)\frac{1}{G\MBH}
\end{align}
where the trace is performed over the late time ``out'' basis, and we defined an entropy production function $\zeta(\MBH, a_\star) = \sum_i \zeta_i(\MBH,a_\star)$, in analogy to the evaporation functions,
\begin{align}
    \zeta_i (M, a_\star) = \frac{g_i}{2\pi} \int_{z_i}^\infty \sum_{l=s_i}\sum_{m=-l}^l \left[(\mathcal{N}_{ilm}+\epsilon_i)\ln(1+\epsilon_i\,\mathcal{N}_{ilm}) - \mathcal{N}_{ilm} \ln \mathcal{N}_{ilm}\right] dx.
\end{align}
\end{widetext}
We computed ${\cal N}_{ilm}$ for 150 values of $x$ in the range $[0.001, 3.0]$ and for 100 values of $\as$ in the range $[0, 0.9999]$. 
For the $l, m$ modes, the spectra were obtained for $m \in [-l, l]$, with $l$ values considered up to 9 for bosons and up to $19/2$ for spin 1/2 fermions.
For Rarita–Schwinger fermions, we followed the procedure in Ref.~\cite{Calza:2024ncn} to compute ${\cal N}_{ilm}$ up to a value of $l= 11/2$.
To validate our numerical computation, we obtained the following values for each spin particle per internal degree of freedom for the entropy production function $\zeta_i$, assuming massless degrees of freedom and a Schwarzschild BH
\begin{align}
 	\zeta_i (M, 0) = 10^{-3}\times \begin{cases}
		3.4648 & \text{for } s_i =0, \\
		1.6855 & \text{for } s_i =1/2, \\
		0.6338 & \text{for } s_i =1, \\
            0.2110 & \text{for } s_i =3/2, \\
		0.0651 & \text{for } s_i =2.
	\end{cases}
\end{align}
We find a $0.00551$\textperthousand, $-0.580$\textperthousand, and $2.17$\textperthousand~difference for $s_i=1/2,1,2$, respectively, with the values presented in Refs.~\cite{Page:1983ug,Page:2004xp,Page:2013dx}.
  
The entropy production ratio $\beta$, defined as~\cite{Zurek:1982zz,Page:1983ug}
\begin{align}
    \beta = \frac{dS_{\rm rad}/dt}{-dS_{\tt BH}/dt},
\end{align}
determines by the amount of entropy in the emitted radiation times the decrease of the Bekenstein-Hawking entropy. 
For BHs heavier than $M \gtrsim 10^{17}~{\rm g}$, emitting only neutrinos, and massless photons and gravitons, we find that $\beta = 1.61862$, a value with a difference of $-48.66$ parts-per-million (ppm) with the value given in Refs.~\cite{Page:1983ug,Page:2004xp,Page:2013dx}.
Neutrino emission is expected to become suppressed when $M\gtrsim 10^{26}~{\rm g}$, assuming the lightest neutrino to have a mass of $0.01~{\rm eV}$~\cite{Lunardini:2019zob}. Thus, for heavier BHs, the entropy production ratio drops to $\beta = 1.48471$, with a difference with the value given in Ref.~\cite{Page:2013dx} of $6.6$ ppm.
For BH masses below $M \lesssim 10^{17}~{\rm g}$, additional SM and potentially beyond-the-SM degrees of freedom are emitted, altering the $\beta$ ratio from the values discussed earlier. 
As an example, for a Schwarzschild BH with an initial mass of $10^9~{\rm g}$, emitting all SM degrees of freedom but no additional particles, we find $\beta = 1.64064$. 

To illustrate how the $\beta$ parameter depends on the set of particles that can be emitted, Fig.~\ref{fig:Page_ratio} shows the {\tt BH}-entropy decrease, $8\pi \varepsilon(M)$, and the radiation entropy generation functions (top panels), as well as the $\beta$ parameter (bottom panels) for a Schwarzschild BH as a function of its instantaneous mass. 
In all panels, we observe a decrease in the value of $\beta$ whenever a mass threshold is crossed. These decreases arise from the differing dependencies of the Bekenstein-Hawking entropy decrease and the radiation entropy production functions on the particle masses, parametrized via $z_i$. Specifically, the growth of $8\pi \varepsilon(M)$ outpaces that of $\zeta(M, a_\star)$, leading to a reduction in $\beta$.

In the right panel of Fig.~\ref{fig:Page_ratio}, we consider the mass range $10^9~{\rm g}\leq M \leq 10^{28}~{\rm g}$, assuming the emission of SM particles along with neutrinos treated as massless particles (orange dashed), Majorana fermions (green solid), and Dirac fermions (blue dotted). For the latter cases, the lightest neutrino mass is set to $m_0 = 0.01~{\rm eV}$, while the other masses 
 are computed taking into account the latest results of the global fit of neutrino oscillation data~\cite{Esteban:2024eli}. 
In the Dirac scenario, it is expected that light sterile right-handed states are also emitted alongside the active neutrinos, adding six additional fermionic degrees of freedom to the evaporation process~\cite{Lunardini:2019zob}.
This modifies the dependence of $\beta$, such that for $10^{17}~\g \lesssim M \lesssim 10^{24}~\g$, range in which only neutrinos, alongside with massless particles, are emitted, the ratio takes a value of $\beta = 1.62811$.

In the left panel of Fig.~\ref{fig:Page_ratio}, we explore the mass range $10~{\rm g} \leq M \leq 10^7~{\rm g}$, considering two illustrative extensions of the SM. The first is a minimal supersymmetric (SUSY) scenario, where the superpartners of the SM degrees of freedom are assumed to be present at a scale of $10^5~{\rm GeV}$ (red dot-dashed line). The second scenario involves a dark sector (DS) with 10 copies of the SM particle set, each with masses of $10^{10}~{\rm GeV}$. For comparison, the SM values are presented in light-blue color.
We observe that $\beta$ changes to 1.72377 for the SUSY scenario when $M\leq 10^2~{\rm g}$, and for the DS scenario we find the same value as in the SM case.
This is because the number of particle species, both bosons and fermions, remains unchanged in the DS scenario. Only the total number of particles differs, resulting in the same ratio as in the SM.

\begin{figure}[t!]
    \centering
        \includegraphics[width=\linewidth]{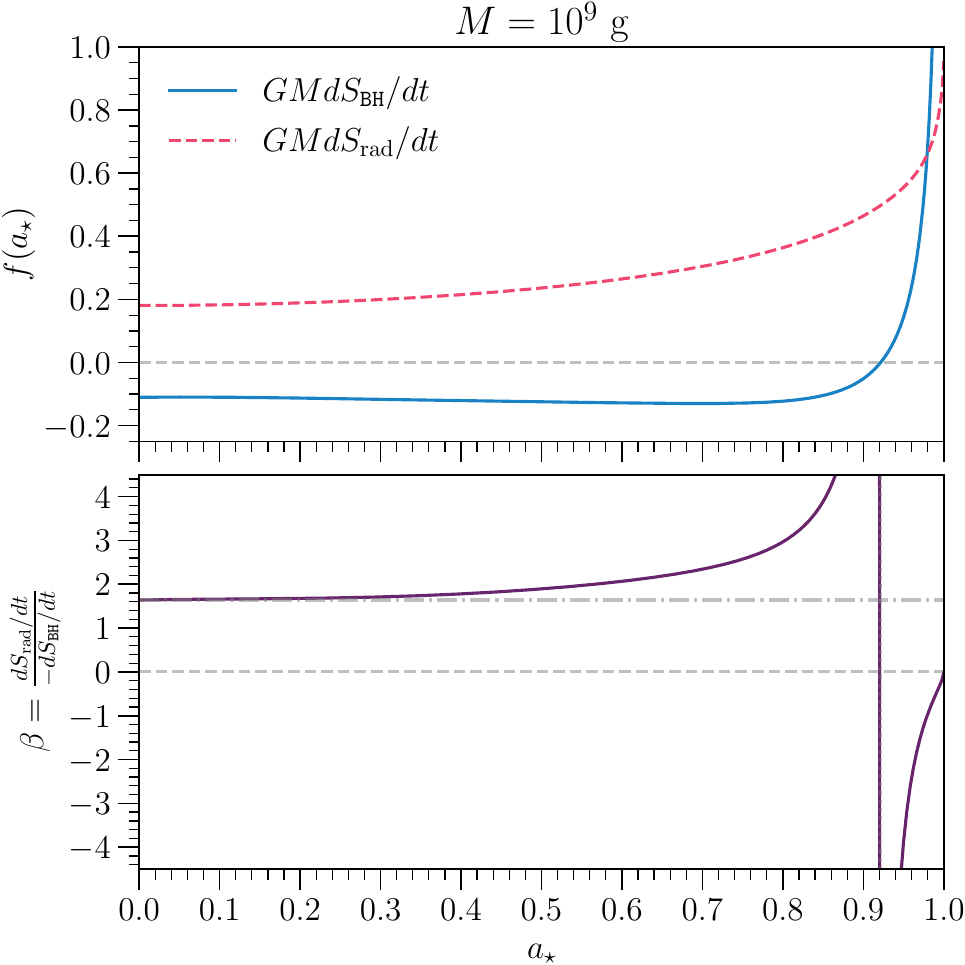}
    \caption{Dimensionless Bekenstein-Hawking entropy decrease rate $GMdS_{\tt BH}/dt$ (solid blue) and von Neumann radiation entropy generation rate $GMdS_{\rm rad}/dt$ (dashed red) (top) and the entropy ratio $\beta$ (bottom) as functions of spin parameter $\as$ for a Kerr BH
    with a mass of $M=10^9~\g$.}
    \label{fig:Page_beta_Kerr}
\end{figure}
Now, turning to Kerr BHs, Fig.~\ref{fig:Page_beta_Kerr} illustrates the entropy ratio $\beta$ for a BH with a mass of $M=10^9~{\rm g}$ as a function of the initial spin parameter $a_\star^{\rm in}$. 
The top panel displays the dimensionless quantities corresponding to the change in Bekenstein-Hawking entropy, $GMdS_{\tt BH}/dt$, (blue curve) and the von Neumann entropy associated with the emitted radiation, $GMdS_{\rm rad}/dt$, (dashed red curve), while the bottom panel shows the $\beta$ parameter. 
A notable distinction from the Schwarzschild case is that, for a BH emitting all SM degrees of freedom, the thermodynamic entropy change rate becomes zero at $a_\star \approx 0.9215$\footnote{This value depends on the set of particles emitted, which is determined by the instantaneous BH temperature. Ref.~\cite{Page:1976ki} reports a value of $a_\star \approx 0.8868$ for a BH emitting two massless neutrinos, photons, and gravitons.}.
Above this threshold, the rate of change becomes positive, signifying an increase in entropy.  
This behavior arises because, for $a_\star \gtrsim 0.9215$, the emission of superradiant modes dominates over thermal emission, which lead to an increase in entropy, as discussed in Ref.~\cite{Page:1976ki}. The $\beta$ parameter captures this transition, becoming negative for $a_\star \gtrsim 0.9215$ and turning positive once the BH spin parameter falls below  such a value.

\subsection{Page Time and Page curves}
\begin{figure*}[t!]
    \centering
        \includegraphics[width=\linewidth]{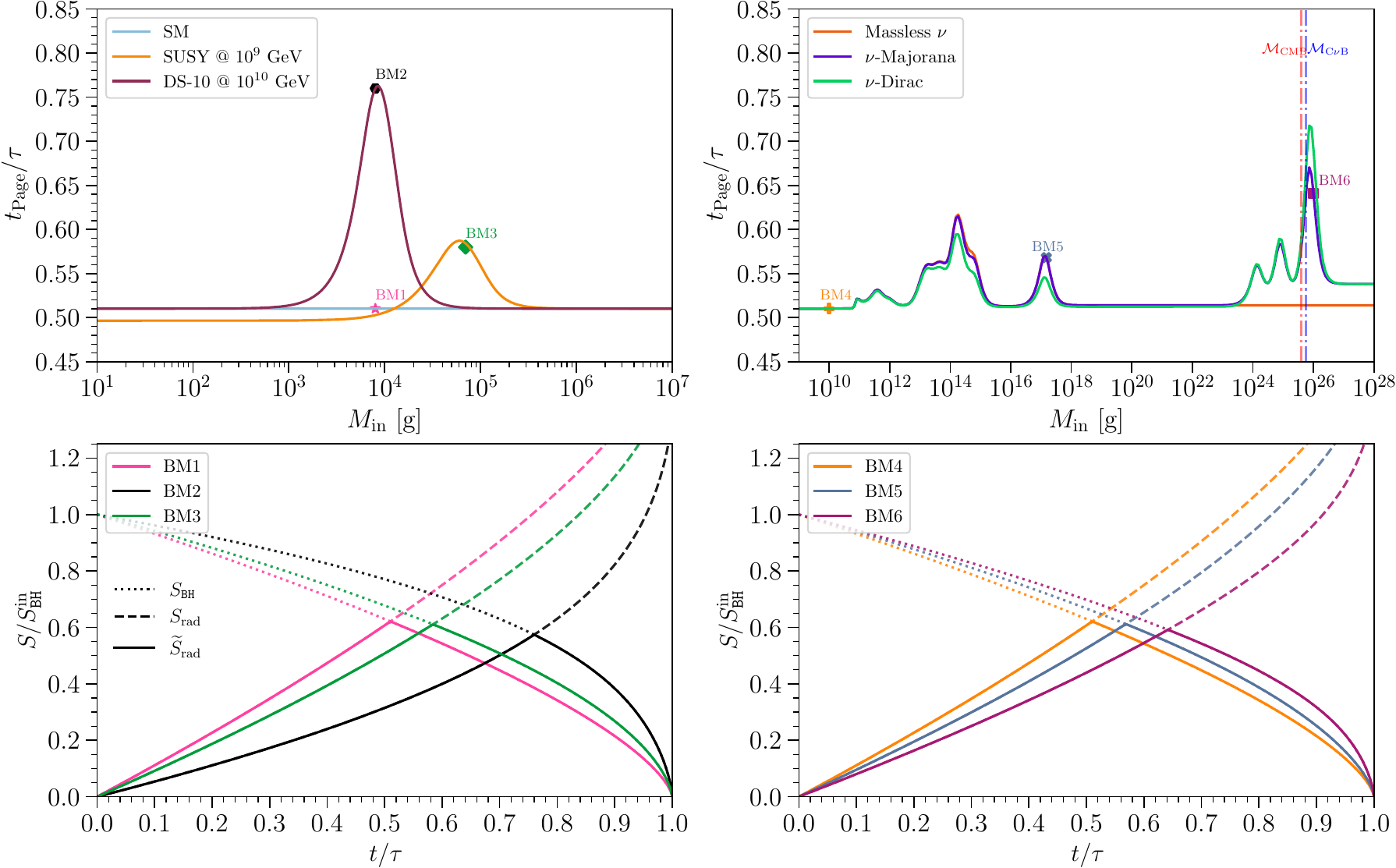}
    \caption{Top panels: Page time normalized to the BH lifetime for the mass range $10~{\rm g} \leq M \leq 10^7~{\rm g}$ (left) and $10^9~{\rm g} \leq M \leq 10^{28}~{\rm g}$ (right) for different sets of existing particles. On the left we assume different particle sets, the SM (light blue), the SM plus a dark sector containing ten copies of it at a scale of $10^{10}~\GeV$ (red), and the minimal supersymmetric extension of the SM at a scale of $10^9~\GeV$. On the right, we show the SM (orange), the SM plus massive Majorana neutrinos (purple), and the SM plus massive Dirac neutrinos (green). Bottom panels: Time evolution of the Bekenstein-Hawking thermodynamic entropy $S_{\tt BH}$ (dotted), emitted radiation entropy $S_{\rm rad}$ (dashed), and the Page curve $\widetilde{S}_{\rm rad}$ (solid) for some chosen benchmark points, cf Tab.~\ref{tab:bench}.}
    \label{fig:Page_curves_Schw}
\end{figure*}
The microscopic interpretation of the Bekenstein-Hawking entropy remains an open question. In ordinary systems, entropy is associated with the number of accessible microstates. Similarly, $S_{\tt BH}$ can be interpreted as quantifying the internal degrees of freedom of a black hole. This idea is encapsulated in the conjecture known as \emph{the Central Dogma}~\cite{Almheiri:2020cfm,Buoninfante:2021ijy}:
\begin{quote}
    A black hole, as observed from the outside, can be characterized as a quantum system with $A/4G$ degrees of freedom, evolving unitarily in time.
\end{quote}
However, if this conjecture holds, we would face the \emph{Information Paradox}.
To address this issue, let us consider a black hole formed from the gravitational collapse of a system initially in a pure quantum state. 
By unitarity, the evolution of the combined system—including the BH and the emitted radiation—must also remain in a pure state.  
However, the emitted Hawking radiation appears thermal and thus in a mixed state, even though the entire system evolves unitarily.  
This apparent contradiction can be understood by viewing Hawking evaporation as the emission of entangled particle pairs: one particle becomes trapped behind the BH horizon, while the other escapes to infinity. Since we can only observe the escaping radiation, the density matrix for the observable system is obtained by tracing over the unobservable BH degrees of freedom. This results in a reduced density matrix that resembles a thermal distribution, representing a mixed state.
Given that the entire system originates from a pure state, the fine-grained entropy of the BH must equal the entropy of the radiation, $S_{\rm rad}$. 

However, as the BH evaporates, its Bekenstein-Hawking entropy decreases. At a specific moment, known as the Page time, $t_{\rm Page}$, these two entropies become equal, $S_{\rm rad} = S_{\tt BH}$.  
Beyond the Page time, the von Neumann entropy of the Hawking radiation exceeds the BH’s remaining degrees of freedom, implying that the combined system of BH and radiation is in a mixed state. 
This leads to the \emph{Information Paradox}~\cite{Hawking:1976ra, Page:1993wv, Almheiri:2020cfm, Buoninfante:2021ijy}.  
To resolve this paradox and recover unitarity, the entropy of the radiation must begin to decrease when $t \sim t_{\rm Page}$~\cite{Page:1993wv, Almheiri:2020cfm,Buoninfante:2021ijy}. 
Therefore, it is conjectured that $S_{\rm rad}$ should follow the evolution of the Bekenstein-Hawking entropy after the Page time.  
In this framework, the actual von Neumann entropy of the Hawking radiation, $\widetilde{S}_{\rm rad}$, is expected to closely follow the minimum between $S_{\rm rad}$ and $S_{\tt BH}$,
\begin{align}
    \widetilde{S}_{\rm rad} = \min(S_{\rm rad}, S_{\tt BH}).
\end{align}  
This characteristic entropy evolution is referred to as the Page curve.

It is important to note that, as a simplifying assumption, we have considered that the BH forms in a pure state. 
More general scenarios allow for the possibility that the BH forms with an initial von Neumann entropy equal to a fraction of its initial Bekenstein-Hawking thermodynamic entropy, entangled with an external reference system~\cite{Page:2013dx,Nian:2019buz}. 
These generalizations lead to differences in the Page curves for the BH and the Hawking radiation, causing their maxima to occur at different times. 
However, for simplicity, we do not explore such scenarios in this work and instead assume that the BH is initially in a pure state.

In the following, we aim to determine how the Page time and the Page curve depend on the BH’s properties—specifically its mass and angular momentum—as well as the spectrum of degrees of freedom present in nature.
We begin by examining Schwarzschild BHs in Fig.~\ref{fig:Page_curves_Schw}. The top panels display the ratio of the Page time to the BH lifetime as a function of the BH mass for the ranges $10~{\rm g} \leq M \leq 10^7~{\rm g}$ (left) and $10^9~{\rm g} \leq M \leq 10^{28}~{\rm g}$ (right). The bottom panels illustrate the time evolution of the Bekenstein-Hawking entropy $S_{\tt BH}$ (dotted line), the  von Neumann entropy of the radiation $S_{\rm rad}$ (dashed line), and the Page curve (solid line) for six benchmark points.

Focusing first on the higher mass range shown in the right panels of Fig.~\ref{fig:Page_curves_Schw}, we find that for a BH emitting only SM degrees of freedom and with $M \lesssim 10^{11}~{\rm g}$, the Page time is $t_{\rm Page} = 0.5102 \, \tau$, where $\tau$ represents the BH lifetime. 
When particle masses start to significantly influence the evaporation process, an intriguing effect emerges: the Page time shifts closer to the lifetime. 
This behavior arises because, depending on the initial BH mass, massive particles may not be significantly emitted during the early stages of evaporation due to Boltzmann suppression. 
However, as the BH temperature increases, the emission of massive particles becomes appreciable once $T \gtrsim m$, where $m$ denotes the particle mass. 
As a result, the BH evaporates more rapidly in the later stages due to the additional degrees of freedom becoming accessible. Moreover, this additional emission of particles leads to an increase in the Hawking radiation entropy, thus leading to a Page time closer to the lifetime.
This effect accounts for the increase in the ratio of the Page time to the lifetime within the range $10^{11}~{\rm g} \lesssim M \lesssim 10^{15}~{\rm g}$, where the masses of quarks and heavy leptons become significant, to be close to $\sim 60\%$ of the BH lifetime. 
The peak in the ratio at $M \sim 1.5 \times 10^{17}~{\rm g}$ is primarily driven by electron-positron emission, while the peaks observed for $M \gtrsim 10^{24}~{\rm g}$ arise from the influence of neutrino masses on their emission. 

A similar behavior is observed for lighter BHs and a few benchmark BSM scenarios, as shown in the left panel of Fig.~\ref{fig:Page_curves_Schw}. 
For a dark sector comprising 10 copies of the SM (dark red curve) at a scale of $10^{10}~{\rm GeV}$, we find a significant increase in the Page time relative to the BH lifetime, reaching $t_{\rm Page} =0.7625\tau$ for a BH mass of $M \sim 10^4~{\rm g}$. 
Similarly, for the SUSY scenario (orange curve), assumed to exist at a scale of $10^9~{\rm GeV}$, we observe an enhancement of the ratio to $t_{\rm Page} = 0.5872\tau$ at a BH mass of approximately $M \sim 7 \times 10^4~{\rm g}$. Notably, for BH masses below $M\lesssim 10^4~\GeV$, the Page time is reduced compared to the SM value, reaching $t_{\rm Page} = 0.4965\tau$. This reduction arises from an increase in the entropy ratio $\beta$, driven by the larger number of scalar degrees of freedom available for emission.
\begin{table}
    \centering
    \begin{tabular}{ccc}
    \toprule\toprule
          & $\MBHi~[\g]$ & Particle set \\ \midrule\midrule
    BM1 & $8\times 10^3$ & SM   \\ \midrule
    BM2 & $8\times 10^3$ & SM + Dark sector - 10 @ $10^{10}~\GeV$   \\ \midrule
    BM3 & $7\times 10^4$ & SM + SUSY @ $10^9~\GeV$   \\ \midrule
    BM4 & $10^{10}$ & SM   \\ \midrule
    BM5 & $10^{17}$ & SM   \\ \midrule
    BM6 & $10^{26}$ & SM + massive neutrinos  \\ \bottomrule
    \end{tabular}
    \caption{Benchmark values for Page curves presented in Fig.~\ref{fig:Page_curves_Schw}.}
    \label{tab:bench}
\end{table}

The bottom panels of Fig.~\ref{fig:Page_curves_Schw} illustrate the evolution of the Bekenstein-Hawking entropy $S_{\tt BH}$ (dotted), the radiation entropy $S_{\rm rad}$ (dashed), and the resulting Page curve (solid), all normalized to the initial value of $S_{\tt BH}$, for selected benchmark points listed in Tab.~\ref{tab:bench}. 
These quantities are plotted as functions of time, normalized to the BH lifetime.  
For all cases presented, a common trend emerges: $S_{\tt BH}$ decreases over time as the black hole loses mass through evaporation, while $S_{\rm rad}$ increases as radiation is emitted and accumulates entropy. 
The Page time varies depending on the specific parameters of each benchmark case.  
This variation is particularly pronounced in the BM2 scenario, which incorporates the dark sector example discussed earlier. 
In this case, the emission of BSM degrees of freedom begins only once the black hole temperature approaches the BSM energy scale. 
Consequently, for the chosen parameters, this additional emission occurs near the end of the black hole’s lifetime, leading to an accelerated decrease in $S_{\tt BH}$ and a corresponding increase in $S_{\rm rad}$. 
As a result, the Page time for BM2 is shifted closer to the black hole’s lifetime, consistent with the behavior explained earlier.  
Similarly, for the right panels, we observe deviations in the entropy evolution driven by the emission of massive particles that become accessible during the later stages of black hole evaporation. 
Specifically, for benchmark cases BM5 and BM6, the emission of electrons and neutrinos becomes significant once the black hole temperature rises to a point where $m/T \sim 1$. 
This late-stage particle emission further impacts the entropy evolution, reflecting the interplay between the Page time and curve and the available degrees of freedom.

\begin{figure}[t!]
    \centering
        \includegraphics[width=\linewidth]{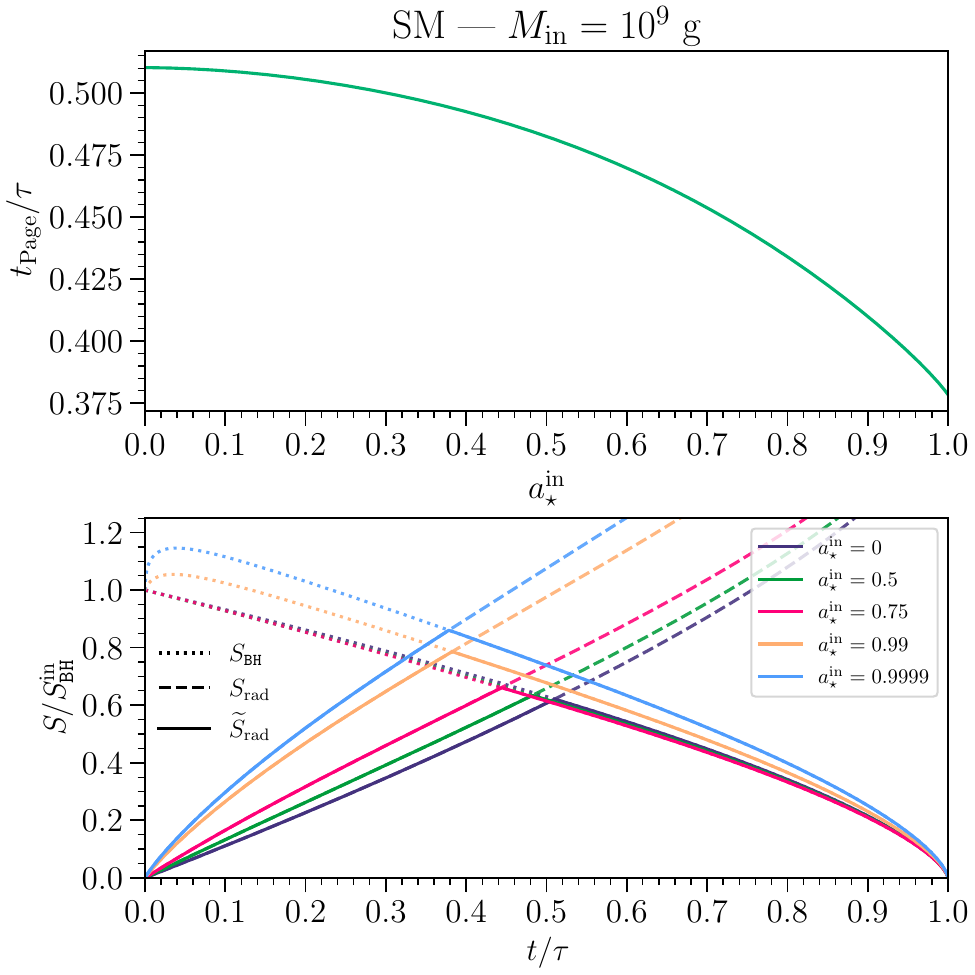}
        \caption{Top: Page time normalized to the BH lifetime as function of the initial spin parameter $\as^{\rm in}$. Bottom: time evolution of the Bekenstein-Hawking thermodynamic entropy $S_{\tt BH}$ (dotted), emitted radiation entropy $S_{\rm rad}$ (dashed), and the Page curve $\widetilde{S}_{\rm rad}$ (solid) for a Kerr BH with $M=10^9~\g$ as function of time normalized to the its lifetime for spin parameters $\as^{\rm in} = \{0, 0.5, 0.75, 0.99, 0.9999\}$.}
    \label{fig:Page_curves_Kerr}
\end{figure}
For Kerr black holes, Fig.~\ref{fig:Page_curves_Kerr} shows the Page time normalized to the black hole's lifetime as a function of the initial spin parameter $\as^{\rm in}$ (top) and the time evolution of the Bekenstein-Hawking and radiation entropies for a black hole with an initial mass of $\MBHi = 10^9~\mathrm{g}$ (bottom) as function of time, normalized to its lifetime. 
In this figure, the black hole emits all Standard Model degrees of freedom plus gravitons, with results presented for different values of $\as^{\rm in} = \{0, 0.5, 0.75, 0.99, 0.9999\}$. 
As expected from the enhanced emission of higher-spin particles by a Kerr black hole, the Page time occurs earlier in the evolution for larger initial values of $\as^{\rm in}$. 
For instance, in the case of $\as^{\rm in} = 0.9999$, we find $t_{\rm Page} = 0.3784\tau$.  
This behavior arises from the distinct evolution of the Bekenstein-Hawking entropy in near-maximally rotating black holes. Early in the evaporation process, the black hole's area, and therefore its thermodynamic entropy, increases due to the dominance of superradiant mode emission, as discussed in Ref.~\cite{Page:1976ki}. This effect is clearly visible in the bottom panel of Fig.~\ref{fig:Page_curves_Kerr} for $\as^{\rm in} = \{0.99, 0.9999\}$.  
Additionally, the enhanced radiation entropy contribution from higher-spin particle emission further shifts the Page time closer to the beginning of the black hole's evolution.

Our results are qualitatively consistent with those of Ref.~\cite{Nian:2019buz}, with two key differences: we account for the full spectrum of SM particles and perform a comprehensive numerical calculation of the greybody factors for a Kerr black hole, rather than relying solely on their low-energy approximations.

Having outlined the general characteristics of the Page time and curve for Kerr black holes, as well as their dependence on the particle content of the Universe, we now turn to examine how these quantities behave for black holes formed in the early Universe, that is, primordial black holes.

\section{Primordial Black Holes}\label{sec:PBHs}

The formation of black holes with masses far below the Chandrasekhar limit --—typical for astrophysical black holes—-- requires large densities, like those that occurred in the Early Universe.
In a radiation-dominated Universe, high density is a necessary but not sufficient condition for black hole formation, as radiation pressure counteracts gravitational collapse. Thus, large density fluctuations are required~\cite{Carr:2020gox,Carr:2020xqk}.  
Several theoretical models have been proposed to explain the origin of these perturbations, which could lead to the formation of primordial black holes (PBHs).
Assuming the existence of an underlying mechanism for BH formation, we can parametrize the resulting PBH population by its initial mass, $\MBHi$, that is assumed to have a monochromatic distribution, and initial energy density, $\rho_{\rm BH}(\Tform)$. In a radiation-dominated Universe, $\MBHi$ is related to the plasma temperature at formation, $\Tform$, as 
\begin{align}
 \MBHi &=\frac{4\pi}{3}\, \alpha\, \frac{\rhorad(\Tform)}{H^3(\Tform)}\sim 10^{14}~{\rm g} \left(\frac{\alpha}{0.2}\right) \left(\frac{10^{10}~{\rm GeV}}{\Tform}\right)^2\,,
\end{align}
where $\alpha \sim 0.2$ is the gravitational collapse factor, and $\rhorad(\Tform)$ and $H(\Tform)$ are the radiation energy density and Hubble rate at formation, respectively.

Constraints arising from Hawking evaporation of PBHs, contingent on when the evaporation occurs, can be summarized as follows.
Assuming only SM degrees of freedom, and that the PBH evolution is given by the semiclassical approximation described above, a black hole with a mass of $M_\star = (5.2898\pm 0.0004) \times 10^{14}~\g$ would have a lifetime equal to the age of the Universe, $t_U = 13.796\pm 0.020~{\rm Gyr}$~\cite{Planck:2018vyg}. 
PBHs with masses in the range $10^8~{\rm g} \lesssim \MBHi \lesssim 2 \times 10^{14}~{\rm g}$ would have fully evaporated either during Big Bang Nucleosynthesis (BBN) or at later times, potentially inducing spectral distortions and modifying the anisotropy power spectrum of the Cosmic Microwave Background (CMB)~\cite{Carr:2020gox,Carr:2020xqk,Keith:2020jww,Acharya:2020jbv,Boccia:2024nly,Altomonte:2025hpt}.
Additional constraints arise from searches for both extragalactic and galactic gamma-ray backgrounds~\cite{Carr:2020gox,Carr:2020xqk,Ballesteros:2019exr,Arbey:2019vqx,Acharya:2020jbv}, imposing strong limits on the mass range $10^{14}~\g \lesssim \MBHi \lesssim 8 \times 10^{14}~\g$.

For heavier PBHs that have not yet fully evaporated but are emitting significant particle fluxes, additional constraints arise from searches for cosmic rays, positrons, and antiprotons. These searches impose limits on the range $10^{15}~\g \lesssim \MBHi \lesssim 10^{17}~\g$~\cite{Boudaud:2018hqb,Dasgupta:2019cae,Laha:2019ssq,DeRocco:2019fjq,Laha:2020vhg,Dutta:2020lqc}.
\begin{figure*}[t!]
    \centering
        \includegraphics[width=0.9\linewidth]{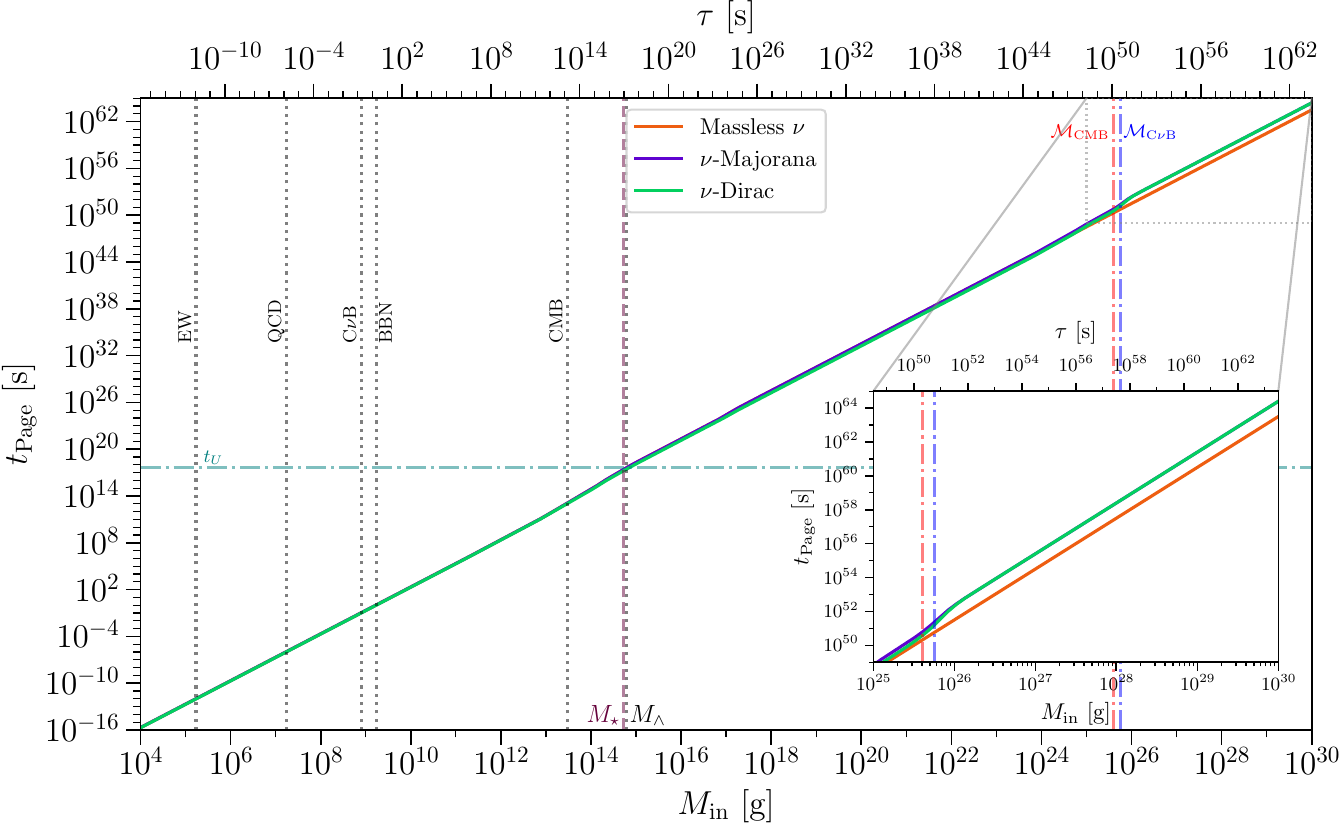}
    \caption{Page time for Schwarzschild BHs as function of its initial mass, assuming SM with massless neutrinos (orange), massive neutrinos assumed to be Majorana (purple) or Dirac (green). The dashed lines indicate PBHs with Page times equal to the electroweak (EW) and QCD phase transitions, neutrino decoupling (labeled as C$\nu$B), BBN and the CMB. $M_\star$ corresponds to a PBH with a lifetime equal to the age of the Universe $t_U$ (shown as a dot-dashed teal line), and $M_\wedge$ to a PBH whose Page time matches the same age.}
    \label{fig:Page_time}
\end{figure*}

We now focus on the determination of the Page time for Schwarzschild PBHs. Figure~\ref{fig:Page_time} shows the Page time as a function of PBH mass, considering three different scenarios: the Standard Model (SM) with massless neutrinos (orange), massive Majorana neutrinos (purple), and Dirac neutrinos (green). 
The vertical lines correspond to PBHs whose Page time coincides with key cosmological eras, including the electroweak (EW) and QCD phase transitions, neutrino decoupling (labeled as C$\nu$B), BBN and the CMB. 
A PBH with a mass of $M_\wedge = (6.2282\pm 0.0004) \times 10^{14}~{\rm g} \approx 1.178~M_\star$ would have a Page time equal to the current age of the Universe. 
PBHs with masses in the range $\MBHi \in [1, 1.178]M_\star = [5.2898, 6.2898] \times 10^{14}~{\rm g}$ would not have fully evaporated yet, but would be in a stage of evaporation where they should follow the Page curve. In this case, the von Neumann entropy of the Hawking radiation would match $\widetilde{S}_{\rm rad}$, assuming the conjecture holds true.

$M_\wedge$ will be modified if the PBH had a significant angular momentum at formation.
Assuming emission of SM degrees of freedoms plus gravitons, we find that $M_\wedge = \{7.297, 8.862, 9.054\}\times 10^{14}~\g$ for PBH which had an initial angular momentum of $\as^{\rm in}=\{0.75, 0.99, 0.9999\}$, respectively.

\subsection{Phenomenological Consequences}

Although it is conjectured that the black hole entropy follows the Page curve, the modification of the particle spectrum from its thermal form after the Page time remains uncertain~\cite{Almheiri:2020cfm}.
While investigating potential deviations from thermality in the Hawking spectra after the Page time lies beyond the scope of this work, one may nonetheless question the possible implications of such modifications on the constraints discussed above, as well as on other aspects of phenomenology, including the generation of dark matter (DM), dark radiation, gravitational waves (GW), and baryon asymmetry (BAU) from PBH evaporation.

As an example, let us consider the production of dark radiation from Kerr PBHs and its contribution to the effective number of neutrino species, parameterized via $\Delta N_{\rm eff}$.
To compute $\Delta N_{\rm eff}$, we adopt the treatment outlined in Refs.~\cite{Hooper:2019gtx,Lunardini:2019zob,Hooper:2020evu,Cheek:2022dbx,Cheek:2022mmy}, to obtain
\begin{widetext}
\begin{align}\label{eq:DNeff}
 \Delta N_{\rm eff} = \left\{\frac{8}{7}\left(\frac{4}{11}\right)^{-\frac{4}{3}}+N_{\rm eff}^{\rm SM}\right\} 
 \frac{\rho_{\rm DR}(T_{\rm ev})}{\rho_{\rm SM}(T_{\rm ev})}
 \left(\frac{g_\star(T_{\rm ev})}{g_\star(T_{\rm eq})}\right)
 \left(\frac{g_{\star S}(T_{\rm eq})}{g_{\star S}(T_{\rm ev})}\right)^{\frac{4}{3}}\,,
\end{align}
\end{widetext}
where $N_{\rm eff}^{\rm SM} = 3.045$ represents the effective number of neutrinos~\cite{deSalas:2016ztq}, $T_{\rm eq} = 0.75~\mathrm{eV}$ denotes the matter-radiation equality temperature, $T_{\rm ev}$ is the plasma temperature at the time of PBH evaporation, $g_\star(T_{\rm ev})$ ($g_{\star S}(T_{\rm ev})$) are the relativistic (entropic) degrees of freedom at evaporation, and $\rho_{\rm SM}(T_{\rm ev})$ and $\rho_{\rm DR}(T_{\rm ev})$ are the energy densities of the SM plasma and dark radiation, respectively, at the same epoch.
As shown in Refs.~\cite{Hooper:2019gtx,Lunardini:2019zob,Hooper:2020evu,Cheek:2022dbx,Cheek:2022mmy}, for PBH masses $M \lesssim 10^9~\mathrm{g}$, $\Delta N_{\rm eff}$ can reach values testable by future CMB surveys. 
In particular, for Kerr black holes, the enhanced graviton emission from highly spinning PBHs can lead to a minimal setup capable of generating $\Delta N_{\rm eff} \sim 0.03$, a value anticipated to be within the sensitivity of upcoming experiments~\cite{Hooper:2020evu,Cheek:2022dbx}.

All these analyses assume the validity of the semiclassical approximation up to scales close to the Planck regime, well beyond the Page time. 
This raises the question of how corrections beyond the Page time might influence such predictions. 
Broadly speaking, two primary effects could impact the determination of $\Delta N_{\rm eff}$: (i) modifications to the PBH lifetime due to unknown phenomena after the Page time, or (ii) changes in the energy density of dark radiation produced by PBH evaporation.

Given the significant uncertainty regarding the precise nature of such modifications, it is instructive to consider instead the amount of dark radiation produced prior to the Page time. 
This is particularly relevant because, if the majority of particles are emitted before the Page time, the determination of $\Delta N_{\rm eff}$ would only be affected if the PBH time evolution were substantially altered after this point. 
Thus, we are naturally led to investigate the fraction of particles emitted before the Page time for Kerr black holes. 

\begin{figure}[t!]
    \centering
        \includegraphics[width=\linewidth]{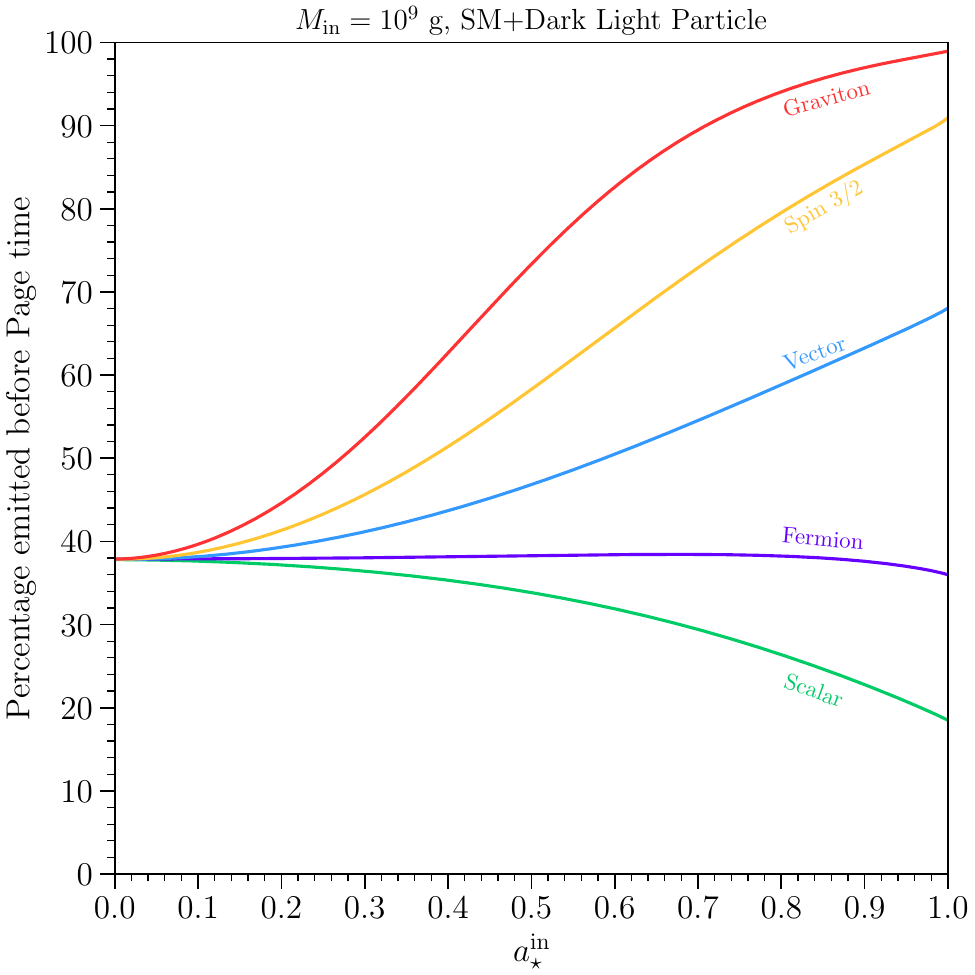}
    \caption{Percentage of particles emitted before the Page time for a Kerr PBH with an initial mass of $\MBHi=10^9~\g$, assuming the emission of all SM degrees of freedom along with a single massless, non-interacting particle. The additional particle is considered to be either a scalar (green), fermion (purple), vector (light blue), spin 3/2 (yellow), or graviton (red).}
    \label{fig:ratio_Kerr}
\end{figure}

Figure~\ref{fig:ratio_Kerr} presents the percentage of particles emitted before the Page time as a function of the initial spin parameter $\as^{\rm in}$ for a PBH with an initial mass of $\MBHi= 10^9~\g$, assuming the emission of SM degrees of freedom plus an additional massless particle, which can be a scalar (green), fermion (purple), vector (light blue), spin 3/2 (yellow), or spin-2 particle (red). 
We find that for Schwarzschild BHs, approximately $37.8\%$ of the total emission occurs before the Page time, regardless of the particle type. 
This result can be understood by noting that the evaporation rate for massless particles remains constant throughout the lifetime of a Schwarzschild BH, even though the absolute emission rate depends on the specific particle type.
Therefore, in this scenario, the contribution to $\Delta N_{\rm eff}$ will be impacted if either the PBH lifetime or the particle production rates are altered after the Page time.

The situation differs for Kerr BHs. As discussed earlier, the emission of higher-spin particles is enhanced for larger values of $\as^{\rm in}$. 
Consequently, for gravitons, we observe that approximately $99\%$ of the total graviton production occurs before the Page time for $\as^{\rm in} \to 1$. 
However, for lower spin values, this percentage decreases, reaching only about $60\%$ for $\as^{\rm in} \sim 0.4$. 
A similar trend is observed for vector and spin 3/2 particles, though their pre-Page-time emission fraction reaches only about $70\%$ and $90\%$ for $\as^{\rm in} \to 1$, respectively. 
In contrast, the emission behavior of fermions and scalars exhibits an opposite trend. For highly spinning BHs, the fraction of emission occurring before the Page time is reduced, with only $\sim 18\%$ of scalars and $\sim 36\%$ of fermions emitted for a nearly maximally rotating PBH.

Consequently, the graviton energy density appearing in Eq.~\eqref{eq:DNeff} is primarily dominated by the well-established semiclassical thermal spectrum. 
However, as demonstrated in Ref.~\cite{Cheek:2022dbx}, dark radiation---specifically in the form of gravitons---experiences additional redshift effects because its emission effectively ceases before the complete evaporation of the PBH. 
Thus, if the PBH evolution deviates from the standard scenario, whether due to changes in the PBH lifetime or modifications in particle emission rates, such deviations would alter the expected dark radiation energy density and, consequently, $\Delta N_{\rm eff}$.
As a result, constraints on PBHs, as well as the generation of DM, BAU, or GWs from evaporation, could be significantly influenced by post-Page-time effects. 

\section{Final Thoughts}\label{sec:Conc}

The information paradox is one of the most significant unresolved problems in theoretical physics. 
Its modern formulation is framed in terms of entropies: assuming that an evaporating black hole forms in a pure state, there exists a moment, known as the Page time, when the semiclassical von Neumann entropy of the emitted Hawking radiation equals the Bekenstein-Hawking entropy, which---according to the \emph{central dogma}---represents the number of internal BH degrees of freedom as seen from the outside. 
After the Page time, the combined black hole and radiation system appears to be in a mixed state, despite having originated from a pure state. 
In a unitary framework, the radiation’s fine-grained entropy should begin to decrease after the Page time, following the  Page curve.

Given its significance, we analyzed how the Page time depends on black hole (BH) properties and the spectrum of emitted particles, considering both the Standard Model (SM), including massive neutrinos, and scenarios beyond it. 
For Schwarzschild BHs emitting all SM degrees of freedom, we found that the Page time occurs at $t_{\rm Page} = 0.5102 \tau$, $\tau$ being the BH lifetime, for initial masses $M \lesssim 10^{11}~{\rm g}$. Once particle masses become important, the Page time shifts closer to the lifetime, as massive particles may not be significantly emitted during the early stages of evaporation due to Boltzmann suppression. When the BH mass decreases, the emission of these massive particles becomes significant, shortening the BH lifetime and increasing the ratio between the Page time and the lifetime. This dependence on particle masses produces peaks in the ratio of the Page time to the lifetime in both SM and beyond-the-SM scenarios.
For Kerr BHs, we found that the Page time occurs earlier for black holes with significant angular momentum. For an initial spin parameter $\as^{\rm in} = 0.9999$, the Page time is $t_{\rm Page} = 0.3784 \tau$, primarily due to the enhanced emission of higher-spin particles at the start of the BH evolution. 

Connecting these results to phenomenology, we determined the Page time for primordial black holes (PBHs) formed in the Early Universe.
A Schwarzschild PBH with an initial mass of $M_\wedge = 6.321\times 10^{14}~{\rm g}$ would have a Page time equal to the age of the Universe, assuming the emission of only SM degrees of freedom. For PBHs with significant angular momentum at formation, $M_\wedge$ increases to $9.059\times 10^{14}~{\rm g}$ for $\as^{\rm in} = 0.9999$. PBHs with masses in the range $\MBHi \in [M_\star, M_\wedge]$, where $M_\star$ represents a PBH with a lifetime equal to the age of the Universe, would already be beyond the Page time but not yet fully evaporated. Therefore, probing this specific, albeit narrow, window in the PBH mass range would be essential for understanding the ultimate fate of the entropy associated with PBHs and the information paradox. 
A direct observation within this range could provide experimental data offering critical insights into these fundamental questions.
In future work, we aim to explore potential avenues for using direct observations of PBHs in this mass range to gain insights into the evolution of radiation and BH entropies beyond the Page time.

Since it is not yet clear how the Hawking spectrum might deviate from its thermal form after the Page time—if the BH entropy indeed follows the Page curve—it remains an open question how such deviations would impact constraints on evaporating black holes or the generation of Dark Matter, Dark Radiation, or baryon asymmetry from PBH evaporation. 
Taking the example of $\Delta N_{\rm eff}$ from graviton emission by Kerr PBHs, we found that most graviton emission occurs before the Page time. Specifically, for PBHs with an initial spin parameter $\as^{\rm in} \gtrsim 0.5$, approximately $75\%$ of the graviton emission takes place prior to this time. 
Therefore, in such cases, the graviton energy density that determines $\Delta N_{\rm eff}$ is dominated by the spectrum computed in the semiclassical approach.

However, the final value of $\Delta N_{\rm eff}$ depends not only on the graviton energy density but also on the temperature of the SM bath at which PBHs evaporate, which is directly tied to their lifetime. 
Any deviations in PBH evolution post-Page-time—whether due to changes in their lifetime or modifications in the particle emission rates—would significantly impact $\Delta N_{\rm eff}$. 
With several proposals supporting the central dogma based on specific models~\cite{Strominger:1996sh,Ashtekar:1997yu,Rovelli:1996dv,Meissner:2004ju,Dvali:2011aa,Dvali:2013eja}, alongside alternative perspectives challenging its validity~\cite{Buoninfante:2021ijy}, the nature of BH evolution beyond the Page time remains an open and unresolved question.
Until a firm consensus on the underlying physics is reached, it is prudent to critically assess the application of models of post-Page-time dynamics within PBH phenomenology, given the current uncertainties.

\section*{Acknowledgments}

I am grateful to Enrico Bertuzzo, Lucien Heurtier, and Jessica Turner for their careful reading of the first version of this work. I also appreciate the warm hospitality of the CERN Theory Group, where part of this research was conducted.
This work was supported by the Consolidaci\'on Investigadora grant CNS2023-144536 from the Spanish Ministerio de Ciencia e Innovaci\'on (MCIN) and by the Spanish Research Agency (Agencia Estatal de Investigaci\'on) through the grant IFT Centro de Excelencia Severo Ochoa No CEX2020-001007-S.

\bibliographystyle{apsrev4-1}
\bibliography{main.bib}

\end{document}